\documentclass[twocolumn,showpacs,pas,prl,english]{revtex4}
\usepackage[T1]{fontenc}
\usepackage[latin1]{inputenc}
\usepackage{graphicx}

\makeatletter



\usepackage{babel}
\makeatother

\begin{document}

\title{Enhanced collimated GeV monoenergetic ion acceleration from a shaped foil target irradiated by a circularly polarized laser pulse}

\author{M. Chen}
\email{mchen@tp1.uni-duesseldorf.de}
\author{A. Pukhov, T.P. Yu}
\affiliation{Institut f\"ur Theoretische Physik I,
Heinrich-Heine-Universit\"at D\"usseldorf, 40225, Germany}
\author{Z.M. Sheng}
\affiliation{Department of Physics, Shanghai Jiao Tong University,
Shanghai 200240, China} \affiliation{Beijing National Laboratory of
Condensed Matter Physics, Institute of Physics, Beijing 100080,
China}

\begin{abstract}
Using multi-dimensional particle-in-cell (PIC) simulations we study
ion acceleration from a foil irradiated by a circularly polarized
laser pulse at 10$^{22}$W/cm$^2$ intensity. When the foil is shaped
initially in the transverse direction to match the laser intensity
profile, the center part of the target can be uniformly accelerated
for a longer time compared to a usual flat target. Target
deformation and undesirable plasma heating are effectively
suppressed. The final energy spectrum of the accelerated ion beam is
improved dramatically. Collimated GeV quasi-mono-energetic ion beams
carrying as much as 18\% of the laser energy are observed in
multi-dimensional simulations. Radiation damping effects are also
checked in the simulations.
\end{abstract}


\maketitle

Ion acceleration by ultraintense ultrashort laser pulse interacting
with solid targets has been extensively studied in the last
decade~\cite{TNSA,shock-acce,Esirkepov2004,other-acce} due to a
number of prospective applications, such as proton
therapy~\cite{therapy}, proton imaging~\cite{ionimaging}, ion beams
ignition for laser fusion~\cite{fusion}, \emph{etc}. Recently along
with the progress of plasma mirror technology, the ion acceleration
from laser-foil interaction has attracted much more attention. It
has been shown in one-dimensional (1D) particle-in-cell (PIC)
simulations that specially by use of circularly polarized (CP) laser
pulses monoenergetic ion beams can be generated in
principle~\cite{Zhang2007,Robinson2008,Klimo2008,Yan2008}. The key
effect here is the suppression of electron
heating~\cite{Zhang2007,Macchi2005}, which otherwise would disperse
the plasma electrons in space and destroy the monoenergetic
acceleration. When a CP pulse is used, the laser ponderomotive force
does not oscillate and mainly pushes the electrons forward. A static
electric field is formed at the laser front. Ions in the region of
electron depletion are then accelerated by this static field.
Theoretical models and simulations based on 1D geometry have shown a
very promising scaling law for the final energy spectrum of the
accelerated ions. It predicts quasi-monoenergetic GeV ion beams for
sufficiently long driver laser pulses
~\cite{Robinson2008,Klimo2008,Yan2008}. This kind of acceleration
belongs to the laser pressure dominated acceleration~(LPDA).

However, multi-dimensional simulations show that the acceleration
structure is not so stable~\cite{Pegoraro2007,Chen2008}. Electrons
and ions are inevitably dispersed transversely when the target is
deformed and heated by the driving pulse. Besides target
deformation, instabilities are also another fatal problem. The
transverse instability of the accelerating structure limits the
maximum energy of the accelerated ions and broadens the final energy
spectrum. To overcome this, we have recently studied the laser mode
effects on the acceleration structure and proposed to use combined
laser pulses for ion acceleration and collimation~\cite{Chen2008}.
In this letter we restudy the problem by considering the target
shaping, which might be easier to realize in experiments. The target
fabrication has already been applied before for the ion acceleration
in the target normal sheath acceleration~(TNSA) regime, where the
optimization of ion energy spectrum can be achieved by target
shaping~\cite{Bulanov2002,Okada2006,Robinson2007}. The experiment
with a micro structured target~\cite{Schwoerer2006} produced a
mono-energetic proton beam in the TNSA regime.

In the LPDA regime, the target has usually a thickness of a few
hundred nanometers. Fortunately, due to the rapid progress in the
nano-technology, structured nano thickness targets can be engineered
today. In the present study, using 2D- and 3D-PIC simulations we
show how to optimize the collimation and mono-chromaticity of the
accelerated ion beams via the target shaping. By optimal matching, a
collimated, GeV quasi-monoenergetic proton beam can be generated by
a CP laser pulse at 10$^{22}$W/cm$^2$ intensity incident on a shaped
foil target~(SFT) with the thickness of a few hundreds nanometer.

\begin{figure}[t]

\begin{centering}\includegraphics[clip,width=75mm]{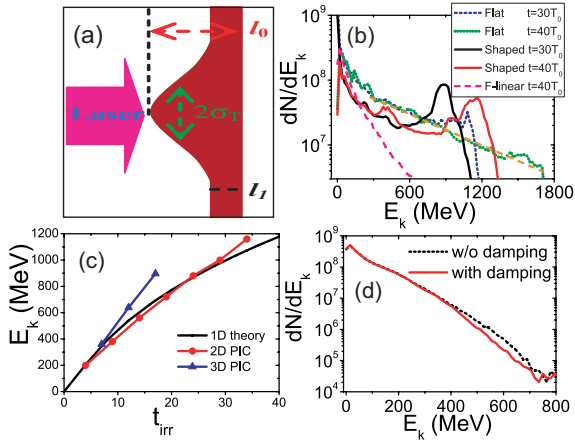} \par\end{centering}
\caption{(color online)\label{layout_spec} (a) Layout of shaped
target. (b) Energy spectrum of ions at $t=30T_0$ and $t=40T_0$. Here
$T_0=3.33fs$ represents the laser period. The orange dashed line
shows the exponential decrease of the spectrum. (c) Energy evolution
of accelerated ions from multi dimensional PIC simulations and 1D
theoretical calculation. Here $t_{irr}$ represents the time of laser
irradiation on target. (d) Energy spectrum of electrons at $t=40T_0$
in the simulations with and without the radiation damping effect.}
\end{figure}

Firstly we study the target deformation under the interaction of a
laser pulse. From the momentum conservation law between the laser
pulse and the target, the evolution of the target area
momentum~($p$) can be described as
following~\cite{Robinson2008,Pegoraro2007}:
\begin{equation}\label{acce}
\frac{dp}{dt} =
\frac{2I}{c}\frac{\sqrt{p^2+\sigma^2c^2}-p}{\sqrt{p^2+\sigma^2c^2}+p},
\end{equation}
where $I$ is the laser intensity, $\sigma$ is the target area
density. For the velocity evolution of the target one obtains:
\begin{equation}\label{velocity}
\frac{d\beta}{dt} =
\frac{1}{2{\pi}n_0m_ic}\frac{E^2(t,x,r)}{l_0}\frac{1}{\gamma^3}\frac{1-\beta}{1+\beta}
\end{equation}
Where $E$ indicates the intensity of laser electric filed, $n_0$ and
$l_0$ are the target initial density and thickness, respectively. It
shows that the energy spread of accelerated ions depends on the
transverse variation of the local ratio of laser intensity to the
target area density. The distance the ions pass in the target is:
$s(r){\propto}{E^2(t,x,r)}{l^{-1}_0}$. An initially flat target is
inevitably deformed, if the laser intensity is not uniform
transversely. The target deformation quickly destroys the
acceleration structure and deteriorates the beam quality. From
Eq.~\ref{velocity} we see that a target can be kept flat if its
areal density $\sigma$ is shaped properly. For the usual
transversely Gaussian pulse, one can use a target with the Gaussian
thickness distribution as shown in Fig.~\ref{layout_spec}(a). In the
following simulations, the distribution of the target thickness
along the transverse direction is:

\begin{equation}\label{thickness}
    l=max\{l_1,l_0{\times}\exp[(-r^2/\sigma^2_T)^m]\}
\end{equation}
Here $r$ is the transverse distance to the laser axis,
$l_1,l_0,\sigma_T, m$ are the shape factors, which are shown in
Fig.~\ref{layout_spec}(a).

First, we use 2D simulations to find the optimal parameter region
because they are computationally less expensive than simulations in
the full 3D geometry. The total simulation box is
$32\lambda(x)\times32\lambda(y)$, here $\lambda$ is the laser
wavelength. The foil plasma consists of two species: electrons and
protons. They are initially located in the region
$5\lambda\leq{x}\leq5.3\lambda$ with the density of $n=100n_c$,
where $n_c=\omega^2m_e/4{\pi}e^2$ is the critical density for the
laser pulse with the frequency $\omega$. For 1${\mu}$m laser pulse
it is $n_c=1.1\times10^{21}/cm^3$. Here, we present the results for
a shaped foil target whose parameters are
$l_0=0.3\lambda,\sigma_T=7\lambda, l_1=0.15\lambda,m=1$. For the
flat target, we just set $l_1=0.3\lambda$, other parameters are the
same. Thus, the total number of ions in the center part of the SFT
is originally less than that in the case of the flat target. The
target is shaped along the Y direction in the 2D case and in the
radial direction in the 3D case. The normalized maximum amplitude of
the laser electric field at the focus is $a_0=eE_0/m{\omega}c=100$.
This corresponds to the laser intensity of
$I=2.76\times10^{22}W/cm^2$ for the assumed wavelength $\lambda=1\mu
m$. The full width half maximum~(FWHM) radius of the focal spot is
$\sigma_L=8\lambda$. The laser pulse has a trapezoidal temporal
intensity profile (linear growth - plateau - linear decrease), with
$1\lambda/c - 8\lambda/c - 1\lambda/c$. Thus, the total laser pulse
energy is about 824J. At $t=0$ the laser pulse enters the simulation
box from the left boundary. Since the normalized laser electric
field we used here is about 100 and the laser pulse is circularly
polarized, we include the electron radiation damping in the VLPL
code and check its effects on the electron
cooling~\cite{Pukhov-vlpl,kiselev-synchrotron}.

\begin{figure}[t]
\begin{centering}\includegraphics[clip,width=75mm]{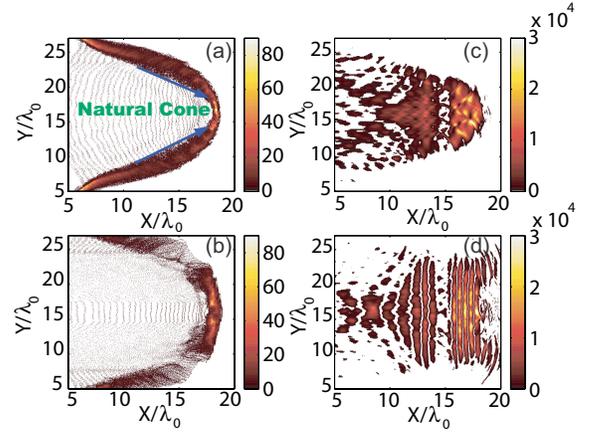} \par\end{centering}
\caption{(color online)\label{density_intensity} Spatial density
distribution of ions in the flat target case (a) and in the SFT case
(b) at $t=25T_0$. Spatial distribution of the laser intensity
($E^2_y+E^2_z$) in the flat target case (c) and in the SFT case (d)
at $t=25T_0$.}
\end{figure}

Fig.~\ref{layout_spec}(b) shows the energy spectrum of the
accelerated ions at $t=30T_0$ and $t=40T_0$ for the flat and shaped
targets in the 2D-PIC simulations. The flat target produces no
obvious peak structure in the spectrum. Instead, the spectrum shows
an exponential decrease like
$dN/dE_k\approx10^{8}\times\exp[-{E_k/E_{\rm eff}}]$ with $E_{\rm
eff}\approx 500$~MeV for $E_k>300MeV$ and a cutoff energy 1.7~GeV at
$40T_0$.

When a SFT is used with the transverse shape factor
$\sigma_T=7\lambda$, the spectrum becomes quasi-monoenergetic. The
energy of the peak is about 880MeV at $t=30T_0$ and later reaches
1.2GeV at $t=40T_0$, which are very closed to the analytical values
obtained by solving Eq.~\ref{velocity}. The analytical values are
shown in Fig.~\ref{layout_spec}(c) by the solid line. As we can see
the maximum ion energy at $t_{sim}=40T_0$ in 2D simulation is a bit
higher than the 1D theoretical value. This is because of reduction
of the target area density during the interaction. Although the
maximum cutoff energy of the ions in the SFT case is lower than that
in the flat target case, much more protons are accelerated in a much
narrower region, which benefits the further application of the
accelerated proton beams.

To show the polarization effect, a linearly polarized laser pulse is
used. The magenta dashed line in Fig.~\ref{layout_spec}(b) shows the
ion energy spectrum at $t=40T_0$. In this case, the electrons are
easily heated and scattered by the oscillating part of the laser
ponderomotive force. The target becomes transparent to the pulse
very soon. Ions are only accelerated by the spatially dispersed
electron cloud and cannot get as high energy as in the CP pulse
case. The spectrum is again exponential with a lower cutoff.

\begin{figure}[b]
\begin{centering}\includegraphics[clip,width=75mm]{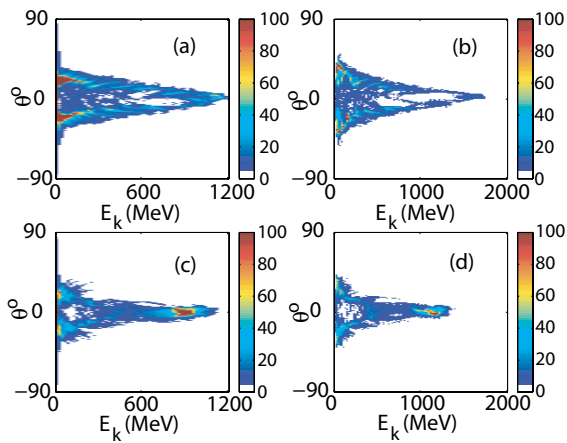} \par\end{centering}
\caption{(color online)\label{angular} Angular distribution of ions
at $t=30T_0$ and $t=40T_0$. (a) and (b) correspond to flat target
case; (c) and (d) correspond to SFT case. The color represents the
relative ion number.}
\end{figure}

Figures \ref{density_intensity}(a,b) show the spatial distribution
of ions at $t=25T_0$ in the two compared simulations. The target
shaping leads to a more transversely uniform ion acceleration. The
initially flat target in contrary is deformed and a natural cone
builds up during the interaction. The laser intensity distribution shown in
Fig.~\ref{density_intensity}(c,d) confirms this. The natural cone
focuses the lateral laser energy to the center and thus reinforces
the on-axis ion acceleration. On one hand, this effects destroys the
foil, but on the other hand it leads to the higher cutoff energy as shown in
Fig.~\ref{layout_spec}(b). Besides this, when the laser pulse
irradiates the cone, electrons are easily extracted by the laser field
out from the inner wall of the cone and heated because of the
oblique incidence. These heated electrons disperse in space and
pollute the acceleration structure, which destroys the
mono-energetic character of the ion spectrum. By use of the shaped
target, these undesirable effects are reduced dramatically.

The angular distributions of the accelerated ions in the two target
cases are presented in Fig.~\ref{angular}. It shows that in the SFT
case the accelerated ions mainly move forward. However, in the flat
target case, only a small portion of the highly energetic ions moves
forward. Ions in the middle energy range get a considerable
transverse momentum. From the simulation, we find the average
emission angle for the ions whose energy is larger than 1~GeV is
about $2.7^\circ$ in the SFT case and $5.22^\circ$ in the flat
target case. The number of ions in this energy range is 1.9 times
larger in the SFT case as compared with the flat target. Clearly,
both the collimation and the total flux of accelerated ions are
improved in the SFT case.

\begin{figure}[t]
\begin{centering}\includegraphics[clip,width=75mm]{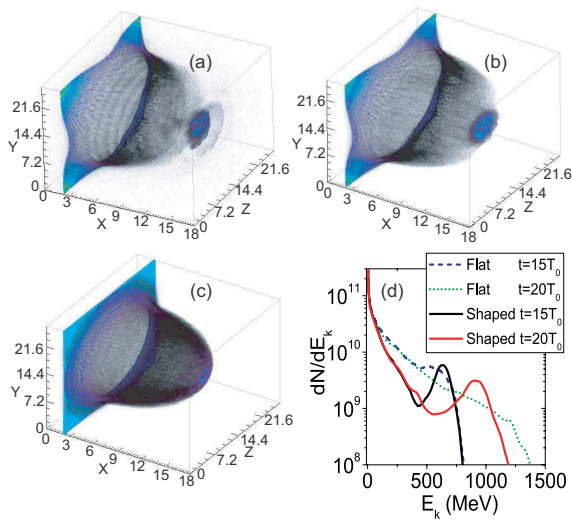} \par\end{centering}
\caption{(color online)\label{3d_sim} Spatial density distribution
of electrons (a) and ions (b) in the 3D simulation at $t=20T_0$. (c)
Spatial density distribution of ions in the flat target case. (d)
Energy spectrum of ions in the SFT case and flat target case at
$t=15T_0$ and $t=20T_0$. At $t=20T_0$ the driven laser pulse has
already left away from the target.}
\end{figure}

To ensure that these effects are not a 2D artefact, we perform full
3D simulations. For the shaped target, we use $\sigma_T=6\lambda$ in
the 3D-simulation. This value appears to be the optimal one for the
assumed laser parameters. The initial position of the target is
moved to $x=2\lambda$ to reduce the computational cost. The laser
longitudinal profile is also reduced to be :$1\lambda/c - 5\lambda/c
- 1\lambda/c$. Other parameters are the same as those in the 2D
simulation above. The electron and ion distributions at $t=20T_0$
are shown in Fig.~\ref{3d_sim}. As we see, in the SFT case a compact
target sheath breaks out from the rest of the foil and is
accelerated by the laser pulse. The same effect was also observed in
the 2D simulation as shown in Fig.~\ref{density_intensity}(b). In
contrast, in the flat target case, Fig.~\ref{3d_sim}(c) displays a
continuously dispersing ion density distribution. The ion energy
spectrum shown in Fig.~\ref{3d_sim}(d) also confirms the
quasi-monoenergetic peak in the SFT case. The number of the ions
with energy larger than 800MeV are $5.09\times10^{11}$ and
$6.63\times10^{11}$ for the flat target and shaped target,
respectively. And their energies are $5.05\times10^{14}$ MeV and
$6.16\times10^{14}$ MeV, the conversion efficiencies are $14.72\%$
and $17.94\%$, respectively. It deserves to note that in the 3D case
the simulation results are far larger than the 1D analytical values
as shown in Fig.~\ref{layout_spec}(c). The calculated peak value of
the ions is 635MeV at $t_{sim}=20T_0$, however, the simulation
result is 910MeV. This larger difference is also due to the target
dispersion, which is stronger in the 3D case. Generally the 1D
estimation gives a higher energy conversion efficiency and a lower
peak energy.

We also change the parameters $l_1$ and $\sigma_T$ to see their
effects on the final spectrum. If we fix other parameters and vary
$l_1$ but keep $l_1<0.2\lambda$, the mono-energetic part of the
final spectrum almost does not change, only the lower energy part
increases with $l_1$. This means that only the ions in the center
part of the target contribute to the final mono-energetic peak.
Correspondingly, the target width $\sigma_T$ is a critical parameter
for the final spectrum. We find in the 2D geometry that when
$\sigma_T/\sigma_L{\approx}0.85$, the optimum spectrum appears. For
the present simulation, when $\sigma_T/\sigma_L\in[0.4,1.0]$ the
monoenergetic peak exists. In the 3D geometry, the optimum value of
$\sigma_T$ is usually smaller than that in the 2D case. Detailed
analysis and simulation work on the target shape parameters such as
$m,l_1,\sigma_T$ is still continuing.

Further, we perform simulations to check the radiation damping
effects, which was found to be very important for transparent nano
targets~\cite{Kulagin2007}. In our simulation parameter
region~($a\leq100$, $n_0=100$ and $l_1,l_0\geq0.1$),
$a=3.33{n_0}{l_0}\approx\pi{n_0}{l_0}$, which satisfies the optimum
matching of target thickness and laser
intensity~\cite{Tripathi2009}, the target is not transparent to the
laser pulse and electrons are not completely exposed to the laser
field. Most of the electrons are inside the target with relatively
low transverse momenta. Our simulations show that only the highest
energy electrons are cooled down as shown in
Fig.~\ref{layout_spec}(d). The x-ray photons are mainly radiated at
the angle of ($\theta=30^0-40^0$) and the average photon energy is
about 10keV. We find that in the present case 21mJ energy is
transferred to the x-ray photons. For the ion acceleration, the very
high energy electrons are not so important, only the bulk of the
ponderomotive energy electrons constructs the local charge
separation field. So in the present case the radiation cooling
effects are not strong enough to benefit ion acceleration. Our
simulation results also demonstrate there is no observable improving
on the final ion spectrum when the radiation damping is considered.

In addition, we should point out that the target shaping only helps
to suppress the target deformation and electron heating. The surface
instabilities still exist. As we can see from
Fig.~\ref{density_intensity}(b) and Fig.~\ref{3d_sim}(c), plasma
bunches have been formed in the plasma front due to these
instabilities. They make a little peak in the ion spectrum in the
earlier periods. However, they evolve further if the laser pulse is
long enough and make electron heating. Suppression of such kinds of
instabilities should be an important work both for the laser ion
acceleration itself and for the fast ignition of inertial fusion
targets based on laser-accelerated ion beams~\cite{fusion}.

In conclusion, by target shaping we have improved both the maximum
peak energy and collimation of the accelerated ions. The shaped
target suppresses the target deformation, electron heating and makes
the ion acceleration much more uniform in the transverse direction
as compared with the plain flat target. Because of the absence of laser
focusing by the natural cone, the maximum cutoff ion energy is smaller
in the shaped target case. However, more ions are concentrated in
the quasi-monoenergetic peak. Radiation cooling effects are checked
and they are not so important in the present laser-plasma
configuration.

This work is supported by the DFG programs TR18 and GRK1203. MC
acknowledges support by the Alexander von Humboldt Foundation. ZMS
is supported in part by the National Nature Science Foundation of
China (Grants No. 10674175, 60621063) and the National Basic
Research Program of China (Grant No. 2007CB815100).

%\bibliography{ionaccepaper}% Produces the bibliography via BibTeX.

\end{document}